\documentclass[12pt]{article}

\usepackage{epsfig}

\usepackage{amssymb}

\def\be{\begin{equation}}
\def\ee{\end{equation}}
\def\ben{\begin{displaymath}}
\def\een{\end{displaymath}}
\def\ba{\begin{array}{c}}
\def\ea{\end{array}}

\newcommand{\ed}{\end{document}}

\begin{document}

\titlepage

\begin{center}{\Large \bf

Asymptotically vanishing ${\cal PT}-$symmetric potentials and
negative-mass Schr\"{o}dinger equations

%
%
%
%

\vspace*{1.2cm}

}\end{center}


\begin{center}

Miloslav Znojil, Petr Siegl  \vspace{3mm}

Nuclear Physics Institute of Academy of Sciences of the Czech
Republic, 250 68 \v{R}e\v{z}, Czech Republic,

\vspace{3mm}

and \vspace{3mm}

  G\'{e}za L\'{e}vai

\vspace{3mm}

Institute of Nuclear Research of the Hungarian Academy of Sciences,

PO Box 51, H-4001 Debrecen, Hungary

\vspace{3mm}

\end{center}

\vspace{5mm}

\vspace{5mm}
\newpage

\section*{Abstract}

In paper I [M. Znojil and G. L\'{e}vai, Phys. Lett. A 271 (2000)
327] we introduced the Coulomb - Kratzer bound-state problem in
its cryptohermitian, ${\cal PT}-$symmetric version. An instability
of the original model is revealed here. A necessary stabilization
is achieved, for almost all couplings, by an unusual, negative
choice of the bare mass in Schr\"{o}diner equation.

\newpage

\newpage

\section{Introduction}

Intuitively one feels that for Schr\"{o}dinger equations
 \be
 \frac{\ \hbar^2}{2m} \,
  \left [-\frac{d^2}{d{x}^2}
  +  \frac{L(L+1)}{{x}^2}\right ]\, \Psi({x})
 +V(x) \, \Psi({x}) = E \,\Psi({x})\,
 \label{SEor}
  \ee
there should exist a close connection between the reality of
potential $V(x)$  and the reality of the corresponding energies
$E$. Unfortunately, such a type of intuition proves deceptive.
Recent studies (e.g., \cite{BB} or \cite{DDT}) showed that many
manifestly non-Hermitian potentials, e.g.,
 \be
  V^{(BB)}(x) = x^{2}\,({\rm i}x)^{4\delta}\,, \ \ \ \ \ \ \delta
  \geq 0\,
  \label{BBmod}
  \ee
still lead to a full reality of the spectrum. The key to such an
unexpected phenomenon can be seen in the Bender's and Boettcher's
\cite{BB} fortunate choice of a complex integration contour
$x=x^{(BB)}(s)$ in eq.~(\ref{SEor}). Its asymptotes
 \be
  x^{(BB)}(s)\ \approx\
  \left \{
  \begin{array}{ll}
  s \,e^{{\rm i} \varphi}\,,
 \ \ \ \ \ &s \gg 1\,,\\
  s \,e^{-{\rm i} \varphi}\,,
 \ \ \ \ \ &s \ll -1\,
 \ea
 \right .
 \label{bc}
 \ee
were restricted to the $\delta-$dependent interval of angles,
 \be
 \varphi + \frac{\pi}{2}
 \in
 \left (
  \frac{\pi}{4+4\delta}, \frac{3\pi}{4+4\delta}
 \right )\,.
 \label{optimal}
 \ee
The curve itself was required not to cross the singularity of
$V^{(BB)}(x)$ in the origin, ${\rm i}x^{(BB)}(0)>0$. In this
setting one can impose the standard Dirichlet boundary conditions
at the ends of the left-right-symmetric curve of complex
coordinates, $\Psi({x^{(BB)}(\pm \infty)})=0$, with the
computationally preferred slope lying precisely in the center of
the interval,
 \be
 \varphi^{(BB)}
 =
  \frac{\pi}{2+2\delta} - \frac{\pi}{2}\,.
  \label{SEBB}
 \ee
In ref.~\cite{BB} it has been emphasized that potentials
(\ref{BBmod}) as well as paths of $x$ and angles (\ref{SEBB}) were
chosen as symmetric with respect to the combination of the
parity-reversal symmetry mediated by the operator ${\cal P}$ with
the time-reversal symmetry represented by operator ${\cal T}$ (cf.
also ref.~\cite{BG} in this respect). In ref.~\cite{BBjmp} it has
been added that for the other eligible domains of angles, say, for
 \be
 \varphi + \frac{\pi}{2}
 \in
 \left (
  \frac{3\pi}{4+4\delta}, \frac{5\pi}{4+4\delta}
 \right )\,
 \label{nextal}
 \ee
the reality of the spectrum breaks down at some non-vanishing
exponents $\delta<\delta_0$. In this sense the specific ${\cal
PT}-$symmetric choice of (\ref{BBmod}) -- (\ref{optimal}) giving
$\delta_0=0$ may be considered optimal.

The discussions in refs. \cite{BB,BBjmp} did not involve the
negative exponents $\delta$ and, in particular, the short-range
models where $V(\infty)=0$. The gap has partially been filled by
ref.~\cite{origi} where we studied eq.~(\ref{SEor}) with one of the
simplest possible asymptotically vanishing ${\cal PT}-$symmetric
potentials of the Coulomb-Kratzer two-parametric form,
 \be
 V(x) = V^{(CK)}(x) =\frac{{\rm i}Z}{x}+\frac{F}{x^2}\,.
 \label{SEck}
 \ee
This model admits
 $
 \varphi
 \in
 \left (
  0,\pi
 \right )$
(cf. eq. (\ref{optimal}) with $2\delta = -1$). From eq.~(\ref{SEBB})
giving $
 \varphi^{(BB)}=
 \pi/2$ one arrives at the U-shaped
complex-coordinate contours $x^{(BB)}(s)$ as sampled in Figure
\ref{obr0ita} where the cut is assumed from $x=0$ upwards.
Marginally let us emphasize that our Schr\"{o}dinger
eq.~(\ref{SEor}) in the most common physical setting using an
integer angular momentum $\ell = 0, 1, \ldots$ should in fact be
considered with the ``centrifugal-like" term of a non-integer
strength $L(L+1) = \ell(\ell+1) + F$ in general.

\begin{figure}[h]                     
\begin{center}                         
\epsfig{file=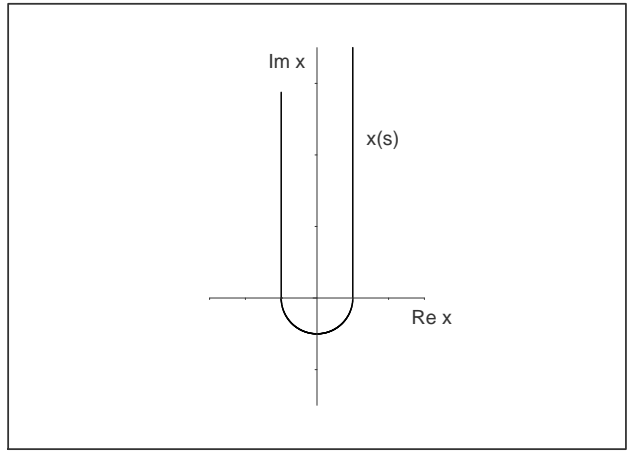,angle=270,width=0.6\textwidth}
\end{center}                         
\vspace{-2mm} \caption{The optimal, U-shaped contour of the
complexified coordinates $x^{(BB)}(s)$ for the Coulomb-Kratzer
${\cal PT}-$symmetric  potential (\ref{SEck}). 
 \label{obr0ita}}
\end{figure}



At the time of the publication of ref.~\cite{origi} (to be cited
as paper I from now on) the physical meaning of the similar models
remained still rather obscure. Many authors studied and
interpreted them as mere effective non-Hermitian simulations of
spectra, not allowing any immediate physical interpretation of the
related wave functions $\Psi(x) \in I\!\!L_2(I\!\!R)$. Although we
also accepted the same philosophy in paper I, we were aware of the
fact that such an attitude significantly weakened the impact and
practical applicability of similar studies.

Fortunately, the subsequent development of the subject clarified
that the potentials as exemplified by eq.~(\ref{SEck}) can be
interpreted as fully compatible with the standard postulates and
probabilistic interpretation of Quantum Mechanics. One of the most
straightforward {\em mathematical} keys to the resolution of such
an apparent puzzle can be seen in the existence of a suitable {\em
non-unitary} invertible map $\Omega$ between some manifestly
non-Hermitian Hamiltonians $H \neq H^\dagger$ and their manifestly
Hermitian partners $\mathfrak{h} = \Omega\,H\,\Omega^{-1}$ (cf.,
e.g., ref.~\cite{SIGMA} for a compact explanation of this
mathematical idea).

From the point of view of physics, the historical origin of the
idea of relevance of isospectrality between $\mathfrak{h}$ and $H$
can be traced back to the study of models of atomic nuclei
\cite{Geyer}. There an explicit example of operator $\Omega \neq
(\Omega^\dagger)^{-1}$ has been provided by the generalized Dyson
mappings \cite{Janssen}. Our recent return to these physical
studies in our mathematical review \cite{SIGMA} showed that for
the ${\cal PT}-$symmetric models all the probabilistic physical
postulates of quantum theory remain valid.

Among immediate and most recent phenomenological applications of
non-Hermitian, ${\cal PT}-$symmetric operators $H\neq H^\dagger$
with real spectra let us mention here just the preprint
~\cite{grav} dealing with a ${\cal PT}-$symmetric version of a
flat Friedmann model in quantum cosmology. In such a broader
physical context we feel particularly inspired here by one of
technical questions discussed in this paper and concerning the
possible instabilities of generic ${\cal PT}-$symmetric systems.
In this sense we also returned to our older results of paper I
which will be re-evaluated, corrected and re-interpreted in what
follows.


\section{Free motion along asymptotes}

In the majority of presentations of Schr\"{o}dinger eq.~(\ref{SEor})
in textbooks  one works with the real $x$ specifying the position of
a particle or quasiparticle which carries a constant mass $m=m_0>0$.
The influence of external forces is modeled solely by a potential
$V(x)$. During the last few years a manifest coordinate-dependence
of the mass term has been allowed as well \cite{tri}. The choice of
$m=m(x)$ opened new perspectives in an optimal description of the
effects of medium.

This idea could easily be transferred to the present class of models
where $x=x(s)$ is complex and where the effect of the potential
becomes negligible in the asymptotic domain of $|s| \gg 1$. In such
a setting the mass can be perceived as a potentially
position-dependent {\em complex} quantity, $m=m[x(s)] \in\,
l\!\!\!C$.

At the large $|s|$ our Hamiltonians get approximated by the
kinetic-energy operator $T$ which, by itself, gets complexified in
the light of eq.~(\ref{bc}),
 \be
 T=-\frac{\ \hbar^2}{2m_0} \,\frac{d^2}{d{x}^2}=
  \left \{
  \begin{array}{ll}
   -\,e^{-2{\rm i} \varphi}\,\frac{\ \hbar^2}{2m_0}
    \,\frac{d^2}{d{s}^2}\, ,
 \ \ \ \ \ &s \gg 1\,,\\
 \\
   -\,e^{+2{\rm i} \varphi}\,\frac{\ \hbar^2}{2m_0}
    \,\frac{d^2}{d{s}^2}\, ,
 \ \ \ \ \ &s \ll -1\,.
 \ea
 \right .
 \label{ubc}
 \ee
Once we introduce the asymptotically constant complex effective
local mass $m_{eff}[x(s)]$ it will only depend on the slope
$\varphi$ and on the sign of $s$,
 \be
 T=-\frac{\hbar^2}{2m_{eff}} \,\frac{d^2}{d{s}^2}\,,
 \ \ \ \ \ \ \
  m_{eff}=m_{eff}[x(s)]=\left \{
  \begin{array}{ll}
   \,e^{2{\rm i} \varphi}\,{m}_0\, ,
 \ \ \ \ \ &s \gg 1\,,\\
 \\
   \,e^{-2{\rm i} \varphi}\,{m}_0\, ,
 \ \ \ \ \ &s \ll -1\,.
 \ea
 \right .
 \label{mubc}
 \ee
This observation is too abstract, for several reasons. First of all,
a subtle balance between the left and right branches of wave
functions $\Psi[x(s)]$ exists and reestablishes the reality of the
energies for numerous complex interactions $V[x(s)]$ \cite{Carl}.
Secondly, for $m=m(x)$ the well-known von Roos' \cite{tri} ambiguity
of the kinetic energy would emerge at the finite values of $s$. For
complex $x(s)$ the manifest introduction of the
coordinate-dependence in the mass might also lead to many other
technical complications. For these reasons our present attention
will solely be paid to the models where $m_{eff}$ remains constant.
Using just the asymptotically vanishing potentials exemplified by
eq.~(\ref{SEck}) and assuming the local reality of the kinetic
energy we shall only make a choice between $\varphi=0$ and
$\varphi=\pi/2$. In this way we encounter either the entirely
traditional textbook straight-line models at $\varphi=0$ or their
U-shaped-line innovations at $\varphi=\pi/2$.

%
\begin{figure}[h]                     
\begin{center}                         
\epsfig{file=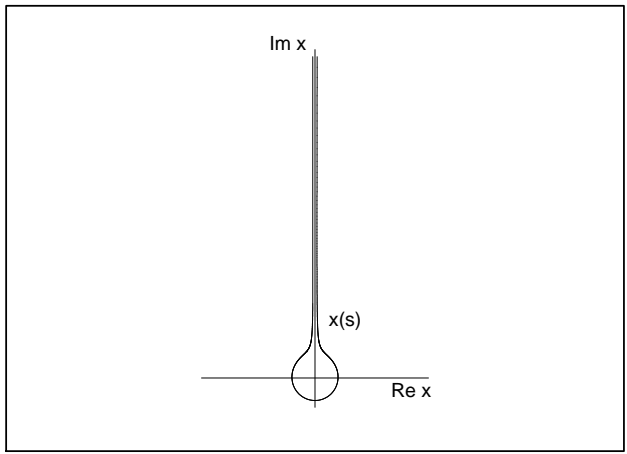,angle=270,width=0.6\textwidth}
\end{center}                         
\vspace{-2mm} \caption{The $\varepsilon=0$ contour
$x^{(U)}(s)=x^{(U)}_{(0)}(s)$.
 \label{obr1a}}
\end{figure}



As long as the former case is very traditional let us only discuss
the choice of $\varphi=\pi/2$ giving the U-shaped contours sampled
in Figure \ref{obr0ita}. Since {\em both} their asymptotes parallel
the {\em upper } imaginary half-axis (i.e., a cut from $x=0$
upwards), the phase of the complex numbers will be assumed lying in
the interval $(-3\pi/2,\pi/2)$. Under such a convention and in terms
of a suitable width parameter $\varepsilon>0$ we may parametrize the
contours of Fig. \ref{obr0ita} as follows,
 \be
 x(s)=x^{(U)}_{(\varepsilon)}(s)\,= \,
 \left \{
 \begin{array}{ll}
 -{\rm i}(s+\frac{\pi}{2}\varepsilon)
 -\varepsilon, & s \in (-\infty, -\frac{\pi}{2}\varepsilon),\\
 \varepsilon e^{ {\rm i}({s/\varepsilon-1/2\pi }
  )},&  s \in
 ( -\frac{\pi}{2}\varepsilon,
 \frac{\pi}{2}\varepsilon),\\
  {\rm i}(s-\frac{\pi}{2}\varepsilon)+\varepsilon\,, & s \in
(\frac{\pi}{2}\varepsilon, \infty).
 \ea
 \right .
 \label{urva}
 \ee
In the complex plane of $x$ the latter curve exhibits the
double-reflection left-right symmetry $x(-s) = -x^*(s)$ which
combines the spatial reflection ${\cal P}$ with the complex
conjugation ${\cal T}$ (let us recollect that ${\cal T}: {\rm i} \to
-{\rm i}$ mimics time-reversal). Our next Figure \ref{obr1a} shows
how such a curve of the complex coordinates could be deformed in the
limit $\varepsilon=0$. It still encircles the origin at a distance
but its asymptotes already strictly coincide with the upper
imaginary half-axis.


Let us emphasize that for $\varphi=\pi/2$ the
coordinate-independence of the effective mass simplifies the
kinetic-energy operator
 \be
  T=-\frac{\hbar^2}{2m_0} \,\frac{d^2}{d{x}^2}=
 +\frac{\hbar^2}{2m_0}\,\frac{d^2}{d{s}^2} \ \ {\rm at}
  \ \ |s| \gg 1\,.
 \label{auhoSEor}
  \ee
Surprisingly enough, it acquires the wrong sign in the sense that
its spectrum becomes unbounded from below at the positive ``bare
mass" $ m_0>0$. This would make the whole system unstable with
respect to small perturbations and, hence, useless for any
phenomenological purposes.

There are hints that also in a field theoretical framework similar
considerations hold concerning negative kinetic energy encountered
during quantization of classical phantom Lagrangians \cite{grav}.
This encourages us to make our argument more quantitative. Let us
recollect the asymptotic form of our Coulomb - Kratzer
Schr\"{o}dinger equation at $|s| \gg 1$,
 \be
  - \,
  \frac{d^2}{d{x}^2}
  \, \Psi({x})
  = \alpha^2\,E \,\Psi({x})\,,\ \ \ \ \ \ \ \ \alpha^2=\frac{2m_0}{\
  \hbar^2}\ >\ 0\,.
 \label{SEoras}
  \ee
Distinguishing between the positive-energy domain ($E = k^2 > 0$,
$k> 0$) and the negative-energy domain ($E = -\kappa^2 < 0$,
$\kappa> 0$) we may employ the general superposition formula
 \ben
 \Psi({x})=C_+\,\Psi_+({x})+C_-\,\Psi_-({x})\,,\ \ \ \
 x=x(s) \sim {\rm i}\,|s|+\ldots
  \een
and insert the pair of linearly independent solutions of
eq.~(\ref{SEoras}),
 \ben
 \Psi_\pm({x})\ \propto \ \left \{
 \ba
 e^{\pm {\rm i}\alpha\,k\,x}\ \sim\
 e^{\mp \alpha\,k\,|s|}
 \,,\ \ \ \ E = k^2 > 0\,,\\
 e^{\pm \,\alpha\,\kappa\,x}\ \sim\
 e^{\pm \,{\rm i}\,\alpha\,\kappa\,|s|}\,,\ \ \ \  E = -\kappa^2 >
 0\,.
 \ea
 \right .
 \een
The upper option proves linked to the asymptotically vanishing bound
states which were constructed in paper I at $E>0$. In parallel, the
lower-line option reveals the admissibility of the free plane-wave
states at all the negative energies. This implies, in a way
unnoticed in paper I, that there exists also a continuous part of
the spectrum which remains unbounded from below.

In the light of the latter semi-intuitive argument our bound-state
model of paper I appears unstable with respect to perturbations
and, hence, deeply unphysical. This forced us to write the present
addendum to paper I showing, in essence, that a complete remedy of
such a very serious shortcoming is unexpectedly easy. Our key idea
is that once we deform the coordinates we must also turn attention
to  the underlying theory  (cf. \cite{SIGMA}) and re-analyze all
the questions of the mathematical consistency of the model.

\section{Amended Coulomb - Kratzer  bound states\label{secIIIc}}

First of all, we must impose the forgotten but essential requirement
of stability, i.e., of the boundedness of the spectrum from below.
In this sense our present main result is that the latter requirement
can be satisfied rather easily. Formally, it appears equivalent to
the reversal of the sign of the bare-mass parameter, $m_0 = -m_1 <
0$. In order to explain this usual amendment of the model let us
first replace eq.~(\ref{SEoras}) by the modified asymptotic equation
 \be
  - \,
  \frac{d^2}{d{x}^2}
  \, \Psi({x})
  = - \beta^2\,E \,\Psi({x})\,,\ \ \ \ \ \ \ \ \beta^2=\frac{2m_1}{\
  \hbar^2}\ >\ 0\,
 \label{SEorbes}
  \ee
which, {\em mutatis mutandis}, implies that
 \ben
 \Psi_\pm({x}) \propto \left \{
 \ba
 e^{\pm \beta\,k\,x}\ \sim\
 e^{\pm  {\rm i} \beta\,k\,|s|}
 \,,\ \ \ \ E = k^2 > 0\,,\\
 e^{\pm  {\rm i}\,\beta\,\kappa\,x}\ \sim\
 e^{\mp\,\beta\,\kappa\,|s|}\,,\ \ \ \  E = -\kappa^2 >
 0\,.
 \ea
 \right .
 \een
Using the same argument as above we deduce that the continuous
spectrum is positive and that the discrete bound-state energy levels
may be expected negative. In this way the structure of the spectrum
of the non-Hermitian Coulomb-Kratzer  model of paper I is thoroughly
modified and made more similar to its well known textbook
Hermitian-Coulomb-Kratzer predecessor.


We saw that the spectrum of our particular illustrative example as
well as of all the similar asymptotically non-interacting models may
be made acceptable, on physical grounds, only if we complement the
complexification of coordinates by the parallel adaptation of the
bare mass. We must keep in mind that even the complexification of
$x(s)$ itself is often perceived as unusual since it causes the
complete loss of the observability of coordinates. This step has
only recently been accepted as an admissible innovative
model-building recipe which characterizes almost all ${\cal
PT}-$symmetric quantum models.

Our present key recommendation of the choice of a negative mass
$m_0$ may look equally  counterintuitive. We believe that it
deserves to be accepted on similar grounds, as a mere very natural
mathematical consequences of the complexification of $x(s)$. Indeed,
the complexification of $x(s)$ immediately implies a breakdown of
the traditional split of the Hamiltonian into its kinetic- and
potential-energy parts so that the switch to the negative value of
the bare mass parameter $m=m_0=-m_1<0$ is in a one-to-one
correspondence with the guarantee of the stability of the system in
question.

Let us return to a constructive demonstration of consistence of the
negative-mass bound-state problem, recollecting first the results of
paper I where the solvable Coulomb-Kratzer potential has been
inserted in the traditional, positive-mass Schr\"{o}dinger equation,
 \be
  \left[-\frac{d^2}{d{t}^2} + \frac{L(L+1)}{{t}^2}
 +i\,\frac{Z}{t} \right]\, \Psi({t}) = \tilde{E} \,\Psi({t})\,.
 \label{SEcold}
  \ee
This equation has only been considered at non-integer $L$ in paper
I. Here, we shall accept the same constraint and assume that
$L\neq 0, \pm 1, \ldots$. This enables us to retype the formula
for the discrete eigenvalues from paper I,
 \be
 \tilde{E}_{\pm n}=\left [\frac{Z}{2L+1\pm (2n+1)}\right ]^2\,,\ \
  \ \ n = 0, 1, \ldots\,.
  \label{eneres}
 \ee
One feels puzzled when seeing that all of these eigenvalues are
positive.  Indeed, the negative bound-state energies would be
generated by the real Coulomb and Coulomb - Kratzer potentials
\cite{Messiah}. 

In the light of our preceding considerations we know that the
spectrum (\ref{eneres}) must be discarded as unstable. This
resolves the latter paradox and, marginally, it also could throw
new light on some recent attempts of using the Coulomb-like
complexified potentials and/or the negative-mass option in
different contexts~\cite{origibe,drumi,habib}. For example,
Mostafazadeh~\cite{priali} noticed that in the latter preprint
\cite{habib} the ${\cal PT}-$symmetry violation caused by the
complex-scaling transformation of $x$ has led to a negative bare
mass but disabled the authors to cope properly with boundary
conditions.  Actually, the similarity transformation expressing
the complex scaling transformation of $x$ violates the ${\cal PT}$
symmetry and either introduces an imaginary part in the energy or
transforms normalizable states into non-normalizable ones and vice
versa \cite{BT}.

The mathematical core of our present proposal is different. In
essence, our present recipe degenerates to the mere change of the
overall sign of the tentative Coulomb Hamiltonian of
ref.~\cite{origi}. The corrected negative-mass version of the
present update of the ${\cal PT}-$symmetric Schr\"{o}dinger
equation (\ref{SEor}) in its Coulomb - Kratzer exemplification
becomes obtainable from eq.~(\ref{SEcold}) by formal substitution
$\tilde{E}\to -E$. Although this implies the reversal of the sign
of $Z$, such a modification of the potential is inessential since
the eigenvalues (\ref{eneres}) themselves are only proportional to
$Z^2$. Our final negative-mass Coulomb - Kratzer Schr\"{o}dinger
equation may be written in the form
 \be
  \left[\frac{d^2}{d{x}^2} - \frac{L(L+1)}{{x}^2}
 -i\,\frac{Z}{x} \right]\, \Psi({x}) = E \,\Psi({x})\,,
 \ \ \ \ \ \ x \in x^{[\pi/2]}_{(\varepsilon)}(s)\,
 \label{SEc}
  \ee
yielding the bound-state-energy formula
 \be
 E={E}_{\pm n}=-\left [\frac{Z}{2L+1\pm (2n+1)}\right ]^2\,,\ \
  \ \ n = 0, 1, \ldots\,.
  \label{eddneres}
 \ee
In Figure \ref{obrmja} this coupling-dependence of the energy
levels is illustrated via the ten lowest bound states.

\begin{figure}[h]                     
\begin{center}                         
\epsfig{file=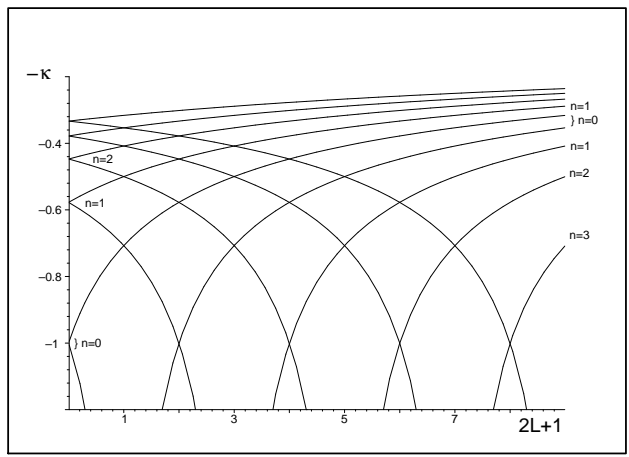,angle=270,width=0.6\textwidth}
\end{center}                         
\vspace{-2mm} \caption{Spectrum of $-\kappa_{n,\sigma}$ at $Z=e=1$
as a function of $2L+1$ with $n = 0, 1, \ldots, 4$ and  $\sigma =
\pm 1$.
 \label{obrmja}}
\end{figure}


Let us summarize that after we changed the sign of the bare mass
the spectrum of our amended ${\cal PT}-$symmetric Coulomb-Kratzer
interaction model looks qualitatively similar to its Coulomb and
Coulomb-Kratzer Hermitian predecessors. Its continuous part of the
spectrum is ``well-behaved" and non-negative, i.e., it is bounded
from below -- this guarantees the stability of the system.
Similarly, all the discrete energy levels only posses  single
accumulation point at $E=0$.

Still, the differences illustrated by Figure \ref{obrmja} are also
worth mentioning (cf., e.g., \cite{Fluegge} for comparison). First
of all, in contrast to the Hermitian Coulomb-Kratzer model the
present discrete spectrum is composed of the two qualitatively
different families of levels which are distinguished by the
$\pm-$ambiguity in formula (\ref{eddneres}). As a consequence, the
traditional ``fall of the particle on the center" known from the
textbooks \cite{Messiah,Fluegge} is now repeated at any integer
``singular value" of our Kratzer-coupling-dependent non-integer
parameters $L = L(F) = 0, 1, \ldots$. This observation also offers
a purely physical explanation why we had to omit these singular
values from our considerations.

In place of the picture we may also employ the following
reparametrization of $2L+1=M_0+\cos^2 \alpha>0$ where the integer
part $M_0\geq 0$ of this parameter is complemented by a small
positive residuum $\cos^2 \alpha <1$ where $\alpha \in (0,\pi/2)$.
This decomposition of  $L=L(M_0,\alpha)$ leads to the
compactification of the ground-state-energy formula
 \be
 E_{(g.s.)}= -\frac{Z^2}{\min (\sin^2\alpha,\cos^2 \alpha)}\,.
 \ee
Although this function of $\alpha$ represents the lower bound of
the whole spectrum, this  function is, by itself, unbounded from
below. This means that in the ``allowed" vicinity of the
``excluded" limiting values of $\alpha=0$ and $\alpha=\pi/2$ our
system still gets very strongly bound. Moreover, even quite far
from $\alpha=0$ and $\alpha=\pi/2$ all the low-lying spectrum
remains extremely sensitive to the small perturbations or
variations of the coupling constant $F$.

\section*{Acknowledgements}
P. S. and M. Z. appreciate the support by the GA\v{C}R grant Nr.
202/07/1307 while G. L. acknowledges the support by the OTKA grant
No. T49646.

\end{document}